\documentstyle[amssymb,preprint,aps]{revtex}
\begin{document}
\title{Microwave-induced $\pi $-junction transition in a \\
superconductor / quantum-dot / superconductor structure}
\author{{Yu Zhu, Wei Li, and }Tsung-han Lin$^{*}$}
\address{{\it State Key Laboratory for Mesoscopic Physics and }\\
{\it Department of Physics, Peking University,}{\small \ }{\it Beijing,}\\
{\it 100871, China}}
\author{Qing-feng Sun}
\address{{\it Center for the Physics of Materials and}\\
{\it Department of Physics, McGill University, Montreal,}\\
{\it PQ, Canada H3A 2T8}}
\date{}
\maketitle

\begin{abstract}
Using the nonequilibrium Green function, we show that microwave irradiation
can reverse the supercurrent flowing through a superconductor / quantum-dot
/ superconductor structure. In contrast with the conventional sideband
effect in normal-metal / quantum-dot / normal-metal junctions, the
photon-assisted structures appear near $E_0=\frac n2\hbar \omega \;(n=\pm
1,\pm 2\cdots )$, where $E_0$ is the resonant energy level of the quantum
dot and $\omega $ is the frequency of microwave field. Each photon-assisted
structure is composed of a negative and a positive peak, with an abrupt jump
from the negative peak to the positive peak around $E_0=\frac n2\hbar \omega 
$. The microwave-induced $\pi $-junction transition is interpreted in the
picture of photon-assisted Andreev bound states, which are formed due to
multiple photon-assisted Andreev reflection between the two superconductors.
Moreover, the main resonance located at $E_0=0$ can also be reversed with
proper microwave strength and frequency.
\end{abstract}

%\vskip 0.4in


\baselineskip 20pt %\baselineskip 12pt

When two superconductors are weakly linked, dc current can flow even without
bias voltage. The driven force of the supercurrent is the phase gradient in
the macroscopic wave function of Cooper pair condensate. The supercurrent
and the phase difference across the junction has the relation $I=I_c\sin
\phi $, with $I_c>0$ being the critical supercurrent. If the link area of
the junction is controlled by some external conditions, $I_c$ may be
enhanced , suppressed, or even reversed. The reverse sign of $I_c$ is
referred as to $\pi $-junction transition, since the minus sign can be
absorbed into $\phi $ as an internal phase shift of $\pi $. The ground state
of $\pi $-junction is $\phi =\pi $ rather than $\phi =0$ as in usual $0$%
-junction. This is easily seen from the relation $I=\frac{2e}\hbar \frac{%
\partial F}{\partial \phi }$ and therefore $F=-\frac \hbar {2e}I_c\cos \phi
+F_0$, where $F$ is the free energy of the junction. The studies of $\pi $%
-junction have not only academic interests, but also potential applications,
e.g., realization of qubit in a superconducting loop with $0$- and $\pi $-
junctions \cite{application}.

In addition to the intrinsic $\pi $ phase shift in the order parameter of
high T$_c$ superconductors due to the d-wave symmetry\cite{HTC}, several
mechanisms were also proposed and investigated for carrying out $\pi $%
-junction with the conventional BCS\ superconductors. (1) {\it Coupling two
superconductors through a ferromagnetic layer}: The idea can be traced back
to the original works of Fulde and Ferrel \cite{FF} and Larkin and
Ovchinnikov \cite{LO}, who independently predicted that superconducting
order parameter can be modulated by an exchange field, and contains nodes
where the internal phase shift is $\pi $. With the growing interests in the
hybrid system of superconducting / ferromagnetic (S/F) materials, $\pi $%
-junction behavior in SFS\ sandwiches and SF superlattices were intensively
studied \cite{sfs1,sfs2,sfs3,sfs4}. Very recently, experiments on Josephson
junction inserted with a weakly ferromagnetic interlayer did observe the $%
\pi $-junction transition \cite{sfsexp}. (2) {\it Coupling two
superconductors by an Anderson impurity or a quantum dot:} The early papers
of Glazman and Matveev \cite{GM} and Spivak and Kivelson \cite{SK} showed
that when the impurity is singly occupied due to Coulomb repulsion, the sign
of Josephson current is opposite to that without the repulsion.. In the last
decade, a number of theoretical papers were devoted to this issue \cite
{sds1,sds2,sds3,sds4}. Experimentally, the technique of fabricating
Josephson junctions containing nanoparticles was available \cite{sdsexp},
yet no relevant results on $\pi $-junction transition was reported. (3) {\it %
Introducing a nonequilibrium distribution in the central mesoscopic region: }%
In a mesoscopic superconductor / normal-metal / superconductor (SNS)
structure, the quasiparticle distribution can be driven far from the
equilibrium by a control voltage across the N region. When the control
voltage exceeds a certain value, the nonequilibrium distribution has so much
weight on the negative part of the current carrying density of states that
the supercurrent reverses its sign. After a few years of the prediction \cite
{sns1,sns2,sns3,sns4}, the reverse of the supercurrent was successfully
observed in a controllable Josephson junction \cite{snsexp}.

In this paper, we propose a new mechanism for the realization of $\pi $%
-junction, schematically shown in the inset of Fig.1. Consider a Josephson
junction consist of two BCS superconductors (S) coupled through a quantum
dot (QD). Applying a microwave (MW) on the QD, the resonant energy level of
QD will shift adiabatically with the time dependent external field. We shall
show below that $\pi $-junction transition occurs in the photon-assisted
Josephson current, and the main resonance can also be reversed by a proper
choice of MW\ strength and frequency. The MW-induced $\pi $-junction
transition is related to the formation of photon-assisted Andreev bound
states (PAABS), which is a generalization of the usual Andreev bound states
(ABS).

We model the S-QD-S system by the following Hamiltonian,

\begin{equation}
H=H_L+H_D+H_R+H_T\;\;,
\end{equation}
where $H_\beta =\sum_{k\sigma }\epsilon _{\beta k}a_{\beta k\sigma
}^{\dagger }a_{\beta k\sigma }+\sum_k\left[ \Delta e^{-\text{i}\phi _\beta
}a_{\beta k\uparrow }^{\dagger }a_{\beta \bar{k}\downarrow }^{\dagger
}+H.c.\right] $ with $\beta =L$ / $R$ is the standard BCS Hamiltonian for
the left / right superconducting\ electrode; $H_D=(E_0+W\cos \omega
t)\sum_\sigma c_\sigma ^{\dagger }c_\sigma $ is for the QD under MW
irradiation, in which the intradot interaction is ignored for simplicity 
\cite{remark}, and $H_T=\sum_{\beta k\sigma }(v_\beta a_{\beta k\sigma
}^{\dagger }c_\sigma +H.c.)$ is the tunnel Hamiltonian , connecting the
three parts together.

The time dependent current flowing through S-QD-S can be formulated using
the nonequilibrium Green function \cite{ar4}, 
\begin{equation}
I_\beta (t)=\frac e\hbar 2%
%TCIMACRO{\func{Re} }
%BeginExpansion
\mathop{\rm Re}%
%EndExpansion
\left\{ Tr\;\sigma _z\int dt_1\left[ G^r(t,t_1)\Sigma _\beta
^{<}(t_1,t)+G^{<}(t,t_1)\Sigma _\beta ^a(t_1,t)\right] \right\} \;,\;
\end{equation}
in which the full Green function of QD and the self energy induced by the
coupling with $\beta $ electrode are defined as

\begin{eqnarray}
G^{r,a,<}(t_1,t_2) &\equiv &\left( 
\begin{array}{ll}
\langle \langle c_{\uparrow }(t_1)|c_{\uparrow }^{\dagger }(t_2)\}\rangle
\rangle & \langle \langle c_{\uparrow }(t_1)|c_{\downarrow }(t_2)\}\rangle
\rangle \\ 
\langle \langle c_{\downarrow }^{\dagger }(t_1)|c_{\uparrow }^{\dagger
}(t_2)\}\rangle \rangle & \langle \langle c_{\downarrow }^{\dagger
}(t_1)|c_{\downarrow }(t_2)\}\rangle \rangle
\end{array}
\right) ^{r,a,<}\;\;, \\
\Sigma _\beta ^{r,a,<}(t_1,t_2) &\equiv &\sum_k\left( 
\begin{array}{cc}
v_\beta ^{*} & 0 \\ 
0 & -v_\beta
\end{array}
\right) g_{\beta k}^{r,a,<}(t_1,t_2)\left( 
\begin{array}{cc}
v_\beta & 0 \\ 
0 & -v_\beta ^{*}
\end{array}
\right) \;\;, \\
g_{\beta k}^{r,a,<}(t_1,t_2) &\equiv &\left( 
\begin{array}{ll}
\langle \langle a_{\beta k\uparrow }(t_1)|a_{\beta k\uparrow }^{\dagger
}(t_2)\rangle \rangle _0 & \langle \langle a_{\beta k\uparrow
}(t_1)|a_{\beta \bar{k}\downarrow }(t_2)\rangle \rangle _0 \\ 
\langle \langle a_{\beta \bar{k}\downarrow }^{\dagger }(t_1)|a_{\beta
k\uparrow }^{\dagger }(t_2)\rangle \rangle _0 & \langle \langle a_{\beta 
\bar{k}\downarrow }^{\dagger }(t_1)|a_{\beta \bar{k}\downarrow }(t_2)\rangle
\rangle _0
\end{array}
\right) ^{r,a,<}
\end{eqnarray}
and $\sigma _z$ in Eq.(2) is the third Pauli matrix, with the diagonal
element +1 for the electron current (spin$\uparrow $) and -1 for the hole
current (spin$\downarrow $). $G^r$ and $G^{<}$ obey the corresponding Dyson
equation and Keldysh equation: 
\begin{eqnarray}
G^r(t_1,t_2) &=&g^r(t_1,t_2)+\int \int dtdt^{^{\prime }}g^r(t_1,t)\Sigma
^r(t,t^{^{\prime }})G^r(t^{^{\prime }},t_2)\;, \\
G^{<}(t_1,t_2) &=&\int \int dtdt^{^{\prime }}G^r(t_1,t)\Sigma
^{<}(t,t^{^{\prime }})G^a(t^{^{\prime }},t_2)\;.
\end{eqnarray}
The remaining task is to solve these equations properly. It should be
pointed out that finite perturbation expansion of Eq.(6) is inadequate in
the problem, because the formation of PAABS involves up to infinite order of
tunneling processes.

Although $G^r(t_1,t_2)$ is no longer the function of $t_1-t_2$, it still
holds that $G^r(t_1+\frac{2\pi }\omega ,t_2+\frac{2\pi }\omega )=G^r(t_1,t_2)
$ due to the periodical time dependence of the MW. Hence $G^r(t_1,t_2)$ can
be Fourier expanded as 
\begin{equation}
G^r(t_1,t_2)=\sum_le^{\text{i}l\omega t_1}\int \frac{d\epsilon }{2\pi }e^{-%
\text{i}\epsilon (t_1-t_2)}\tilde{G}_l^r(\epsilon )\;.
\end{equation}
Define the Fourier transformed Green function 
\begin{equation}
{\bf G}_{mn}^r(\epsilon )=\tilde{G}_{m-n}^r(\epsilon -n\omega )\;\;,
\end{equation}
which has the property that if $C(t_1,t_2)=\int dtA(t_1,t)B(t,t_2)$ then $%
{\bf C}_{mn}(\epsilon )=\sum_k{\bf A}_{mk}(\epsilon ){\bf B}_{kn}(\epsilon )$%
. The Fourier transformed ${\bf g}^r$and ${\bf \Sigma }^{r,<}$ can be
obtained as 
\begin{eqnarray}
{\bf g}_{mn}^r(\epsilon ) &=&\left( 
\begin{array}{cc}
\sum_lJ_{l-m}(\alpha )\frac 1{\epsilon -l\omega -E_0+i0^{+}}J_{l-n}(\alpha )
& 0 \\ 
0 & \sum_{l^{^{\prime }}}J_{m-l^{^{\prime }}}(\alpha )\frac 1{\epsilon
-l^{^{\prime }}\omega +E_0+i0^{+}}J_{n-l^{^{\prime }}}(\alpha )
\end{array}
\right) \;\;, \\
{\bf \Sigma }_{mn}^{r,<}(\epsilon ) &=&\Sigma ^{r,<}(\epsilon -m\omega
)\delta _{mn}\;\;,
\end{eqnarray}
in which $J_n(\alpha )$ is the nth Bessel function with the argument $\alpha
\equiv \frac W{\hbar \omega }$, $\Sigma ^{<}(\epsilon )=f(\epsilon )[\Sigma
^a(\epsilon )-\Sigma ^r(\epsilon )]$ with $f(\epsilon )$ being the Fermi
function, and $\Sigma ^r(\epsilon )=\Sigma _L^r(\epsilon )+\Sigma
_R^r(\epsilon )$ with 
\begin{equation}
\Sigma _\beta ^r(\epsilon )=-\frac{\text{i}}2\Gamma _\beta \frac{\epsilon +%
\text{i}\eta }{\sqrt{(\epsilon +\text{i}\eta )^2-\Delta ^2}}\left( 
\begin{array}{cc}
1 & \frac{-\Delta }{\epsilon +\text{i}\eta }e^{-\text{i}\phi _\beta } \\ 
\frac{-\Delta }{\epsilon +\text{i}\eta }e^{+\text{i}\phi _\beta } & 1
\end{array}
\right) \;\;\;\;\;(%
%TCIMACRO{\func{Im}}
%BeginExpansion
\mathop{\rm Im}%
%EndExpansion
\sqrt{x}>0)\;.
\end{equation}
In the self energy, $\Gamma _\beta $ is the coupling strength between QD and 
$\beta $ electrode, i$\eta $ is the dephasing rate in the S electrodes which
determines the broadening of PAABS. In the numerical calculation, the limit $%
\eta \rightarrow 0^{+}$ is achieved by extrapolation.

The Fourier transformed Dyson equation can be expanded as 
\begin{equation}
{\bf G}^r(\epsilon )={\bf g}^r(\epsilon )+{\bf g}^r(\epsilon ){\bf \Sigma }%
^r(\epsilon ){\bf g}^r(\epsilon )+{\bf g}^r(\epsilon ){\bf \Sigma }%
^r(\epsilon ){\bf g}^r(\epsilon ){\bf \Sigma }^r(\epsilon ){\bf g}%
^r(\epsilon )+\cdots \;\;.
\end{equation}
To re-sum up the series, we adopt the resonant approximation: \cite{ar2} 
\begin{eqnarray}
\frac 1{(\epsilon -l_1\omega -E_0+\text{i}0^{+})}\cdot \frac 1{(\epsilon
-l_2\omega -E_0+\text{i}0^{+})} &\thickapprox &\delta _{l_1l_2}\frac 1{%
(\epsilon -l_1\omega -E_0+\text{i}0^{+})^2}\;\;, \\
\frac 1{(\epsilon -l_1^{^{\prime }}\omega +E_0+\text{i}0^{+})}\cdot \frac 1{%
(\epsilon -l_2^{^{\prime }}\omega +E_0+\text{i}0^{+})} &\thickapprox &\delta
_{l_1^{^{\prime }}l_2^{^{\prime }}}\frac 1{(\epsilon -l_1^{^{\prime }}\omega
+E_0+\text{i}0^{+})^2}\;\;,
\end{eqnarray}
which is justified for the weak coupling case of $\Gamma _\beta \ll \omega $%
. Under this approximation, the Dyson equation can be exactly solved as 
\begin{equation}
{\bf G}_{mn}^r(\epsilon )=\sum_{l\;l^{^{\prime }}}\left( 
\begin{array}{cc}
J_{l-m}(\alpha ) & 0 \\ 
0 & J_{m-l^{^{\prime }}}(\alpha )
\end{array}
\right) \tilde{G}_{ll^{^{\prime }}}^r(\epsilon )\left( 
\begin{array}{cc}
J_{l-n}(\alpha ) & 0 \\ 
0 & J_{n-l^{^{\prime }}}(\alpha )
\end{array}
\right) \;\;,
\end{equation}
with 
\begin{eqnarray}
\tilde{G}_{ll^{^{\prime }}}^r(\epsilon ) &=&\left[ \left( g_{ll^{^{\prime
}}}^r(\epsilon )\right) ^{-1}-\Sigma _{ll^{^{\prime }}}^r(\epsilon )\right]
^{-1}+(-1+\delta _{ll^{^{\prime }}})\left[ \left( g_{ll^{^{\prime
}}}^r(\epsilon )\right) ^{-1}-\tilde{\Sigma}_{ll^{^{\prime }}}^r(\epsilon
)\right] ^{-1}\;\;, \\
g_{ll^{^{\prime }}}^r(\epsilon ) &=&\left( 
\begin{array}{cc}
\frac 1{\epsilon -l\omega -E_0+\text{i}0^{+}} & 0 \\ 
0 & \frac 1{\epsilon -l^{^{\prime }}\omega +E_0+\text{i}0^{+}}
\end{array}
\right) \;\;, \\
\Sigma _{ll^{^{\prime }}}^r(\epsilon ) &=&\sum_k\left( 
\begin{array}{cc}
J_{l-k}(\alpha ) & 0 \\ 
0 & J_{k-l^{^{\prime }}}(\alpha )
\end{array}
\right) \Sigma ^r(\epsilon -k\omega )\left( 
\begin{array}{cc}
J_{l-k}(\alpha ) & 0 \\ 
0 & J_{k-l^{^{\prime }}}(\alpha )
\end{array}
\right) \;\;, \\
\tilde{\Sigma}_{ll^{^{\prime }}}^r(\epsilon ) &=&\left( 
\begin{array}{cc}
\Sigma _{ll^{^{\prime }},11}^r(\epsilon ) & 0 \\ 
0 & \Sigma _{ll^{^{\prime }},22}^r(\epsilon )
\end{array}
\right) \;\;.
\end{eqnarray}

The time dependent current is Fourier transformed as 
\begin{eqnarray}
I_\beta (t) &=&\frac e\hbar \sum_le^{\text{i}l\omega t}I_\beta ^{(l)}\;\;, \\
I_\beta ^{(l)} &=&\frac e\hbar \int \frac{d\epsilon }{2\pi }2%
%TCIMACRO{\func{Re}}
%BeginExpansion
\mathop{\rm Re}%
%EndExpansion
\;Tr\;\sigma _z\left[ {\bf G}^r(\epsilon )\Sigma _\beta ^{<}(\epsilon )+{\bf %
G}^{<}(\epsilon )\Sigma _\beta ^a(\epsilon )\right] _{l0}\;\;.
\end{eqnarray}
With the relations ${\bf G}^{<}(\epsilon )={\bf G}^r(\epsilon )\Sigma
^{<}(\epsilon ){\bf G}^a(\epsilon )$, ${\bf G}^a(\epsilon )=\left[ {\bf G}%
^r(\epsilon )\right] ^{\dagger }$, and ${\bf G}^r(\epsilon )$ solved in
Eq.(16), the current can be evaluated. In this work, we concentrate on the
time averaged current $\bar{I}\equiv I_L^{(0)}=-I_R^{(0)}$. Notice that $%
\bar{I}=\left( \Gamma _RI_L^{(0)}-\Gamma _LI_R^{(0)}\right) /\left( \Gamma
_L+\Gamma _R\right) $, and define $\bar{\Sigma}=\left( \Gamma _R\Sigma
_L-\Gamma _L\Sigma _R\right) /\left( \Gamma _L+\Gamma _R\right) $. Again, we
adopt the resonant approximation in ${\bf G}^{<}(\epsilon )={\bf G}%
^r(\epsilon )\Sigma ^{<}(\epsilon ){\bf G}^a(\epsilon )$, and the current
formula is simplified as 
\begin{equation}
\bar{I}=\frac e\hbar \int \frac{d\epsilon }{2\pi }\sum_{l\;l^{^{\prime
}}}\;J_l(\alpha )\;J_{-l^{^{\prime }}}(\alpha )\;2%
%TCIMACRO{\func{Re}}
%BeginExpansion
\mathop{\rm Re}%
%EndExpansion
\;Tr\;\sigma _z\left[ G_{ll^{^{\prime }}}^r(\epsilon )\bar{\Sigma}%
^{<}(\epsilon )+G_{ll^{^{\prime }}}^{<}(\epsilon )\bar{\Sigma}^a(\epsilon
)\right] \;\;.
\end{equation}
where $G_{ll^{^{\prime }}}^r(\epsilon )=\left[ \left( g_{ll^{^{\prime
}}}^r(\epsilon )\right) ^{-1}-\Sigma _{ll^{^{\prime }}}^r(\epsilon )\right]
^{-1}$and $G_{ll^{^{\prime }}}^{<}(\epsilon )=$ $G_{ll^{^{\prime
}}}^r(\epsilon )\Sigma _{ll^{^{\prime }}}^{<}(\epsilon )G_{ll^{^{\prime
}}}^a(\epsilon )$. One can see in the formula that $\bar{I}$ is contributed
by various photon-assisted Andreev reflections, in which an electron absorbs 
$l$ photons and a hole emits $l^{^{\prime }}$ photons. Eq.(23) is the
central result of this paper.

Below we shall numerically study the influence of the MW field on the
Josephson current flowing through S-QD-S. In the calculation, we set $%
e=\hbar =1$, take $k_BT=0$, and choose $\Delta =1$ as the energy scale. In a
qualitative discussion, we regard $I_c\equiv \bar{I}(\phi =\frac \pi 2)$ as
the critical supercurrent, and use the sign of $I_c$ as the criterion of the
0-junction vs the $\pi $-junction. Fig.1 shows the curves of the critical
supercurrent $I_c$ vs the resonant energy level $E_0$, for different MW\
strengths. There are two remarkable features in the plot: (1) With the
increase of MW strength, the main resonance located at $E_0=0$ is gradually
suppressed, while several photon-assisted tunneling structures grow up
around $E_0=\frac 12n\hbar \omega \;(n=\pm 1,\pm 2\cdots )$. (2) Each of the
structures contains a negative and a positive peak, and $I_c$ reverses its
sign abruptly near $E_0=\frac 12n\hbar \omega $. Meanwhile, the magnitudes
of these structures exhibit a non-monotonous dependence on the MW\ strength.
Feature (1) is quite different from the sideband effect in N-QD-N, where
photon-assisted peaks appear near $E_0=n\hbar \omega $. In N-QD-S junctions,
however, photon-assisted peaks also appear near $E_0=\frac 12n\hbar \omega $%
, which can be explained in the picture of photon-assisted Andreev
reflection (PAAR) \cite{ar2}. Basically, feature (1) can be understood in
the similar way as in N-QD-S, but should be noticed that in S-QD-S the
photon-assisted electrons and holes undergo up to infinite order of PAAR
processes, so that PAABS are formed in the QD (schematically shown in the
inset of Fig.1). Feature (2) is dramatically distinguished from those in
either N-QD-N or N-QD-S, indicating that MW irradiation can lead to $\pi $%
-junction transition in the photon-assisted current through S-QD-S.

To understand the feature (2) requires the knowledge of PAABS . For this
purpose, the time averaged current is rewritten in the form $\bar{I}=\frac e%
\hbar \int \frac{d\epsilon }{2\pi }f(\epsilon )j(\epsilon )$. The analysis
of the current carrying spectrum $j(\epsilon )$ provides the information of
the supercurrent carried by each of the PAABS. Fig.2 illustrates the
spectrum of $j(\epsilon )$ for MW strength $\alpha =0.6$, with $%
E_0=0,\;0.05,\;0.10,\;0.15,\;0.20$. For comparison, the spectrum of $E_0=0$
and $E_0=0.2$ without MW irradiation are also shown on top of Fig.2. Without
MW irradiation, the spectrum of S-QD-S has two ABS within the
superconducting gap, located near $\epsilon =\pm E_0$, carrying supercurrent
with opposite signs. The spectrum also has continuous parts outside the
superconducting gap, negative for $\omega <-\Delta $ and positive for $%
\omega >\Delta $. Turning on the MW, the original two ABS split into two
sets of PAABS, due to various photon-assisted processes. The PAABS marked
with A$_n$ locate near $\epsilon =E_0+n\omega $, carrying supercurrent with
positive signs; while those marked with B$_n$ locate near $\epsilon
=-E_0+n\omega $, carrying supercurrent with negative signs. On the contrast,
MW\ has negligible effect on the continuous spectrum, because the
contribution from these parts depends strongly on the density of states in
the S electrodes and hence photon-assisted processes is largely suppressed.
Notice that at zero temperature $\bar{I}$ is contributed by the spectrum in
the range of $\epsilon <0$. For $E_0=0$ (Fig.2a), the states A$_n$ and B$_n$
are energy degenerate, carrying opposite supercurrent weighted by a
distribution factor $J_n^2(\alpha )$, similar to the sideband in N-QD-N. The
contributions from A$_{-1}$ and B$_{-1}$, A$_{-2}$ and B$_{-2}$, etc. cancel
with each other exactly, and the positive contribution from A$_0$ overwhelms
the negative contribution from C$^{-}$, resulting in a positive current.
With the change of the resonant energy level $E_0$, A$_n$ and B$_n$ move
toward $-\Delta $ and $+\Delta $, respectively. For $E_0=\frac 12\hbar
\omega $ (Fig.2d), the states A$_n$ and B$_{n-1}$ are energy degenerate, and
therefore strongly hybridized with each other. So the weight of A$_n$ and B$%
_{n-1}$ are re-distributed, such that they carry the same amount of
supercurrent with opposite signs. The contributions from A$_0$ and B$_{-1}$,
A$_{-1}$ and B$_{-2}$, etc. cancel with each other exactly, leaving the
negative continuous spectrum C$^{-}$ contribute to the supercurrent. This is
the origin of the $\pi $-junction transition in the photon-assisted
supercurrent. Further investigation shows that the evolution of PAABS\ from
(a) to (d) is more complex than stated above. In fact, the hybridization of A%
$_n$ and B$_{n-1}$ already occurs before the energy degeneracy (see Fig.2d
and its inset). The hybridizing of PAABS obeys the following rule: when two
states A$_i$ and B$_j$ approaches each other, they ``interact'' with each
other only if the energy of A$_i$ is higher than that of B$_j$. With this
rule, the abrupt sign change in the current near $E_0=\frac 12\hbar \omega $
is readily understood.

In the above, we have shown that MW irradiation can reverse the sign of
photon-assisted Josephson current. We are still curious whether the main
resonance located at $E_0=0$ can also be reversed. In the case of $E_0=0$,
the current formula is further simplified since $l=l^{^{\prime }}$ is
required by the resonant tunneling. Fig.3 shows the curves of $I_0\equiv 
\bar{I}(\phi =\frac \pi 2,E_0=0)$ vs the MW\ strength $\alpha $, for four
groups of MW frequencies. These numerical results reveal some of the
properties of $I_0$: (1) $I_0$ has non-monotonous dependence on $\alpha $,
which is natural due to the oscillatory nature of Bessel function. (2) For
some frequencies, e.g., $\omega =0.3$ in Fig.1, $I_0$ is always positive;
while for others, e.g., $\omega =0.6$, $I_0$ reverses its sign in a certain
range of $\alpha $. In other words, MW\ with proper strength and frequency
may also induce $\pi $-junction transition on the main resonance. (3) The
dependence of $I_0$ on $\alpha $ changes abruptly around the frequency of $%
\omega =\Delta /n$ $(n=1,2,3\cdots )$. Roughly, the reason is that when the
frequency crosses $\Delta /n$, A$_{\pm n}$ and B$_{\pm n}$ jump out of the
superconducting gap and feel the singularity in the density of states of S\
electrodes.

In conclusion, we have shown theoretically that MW irradiation has a
non-trivial effect on the Josephson current flowing through S-QD-S
structure.. The photon-assisted tunneling structures appear near $E_0=\frac n%
2\hbar \omega $, and each contains a negative and a positive peak,
indicating $\pi $-junction transition induced by MW. The main resonance
located at $E_0=0$ may also reversed with proper MW strength and frequency.
The understanding of these results requires the knowledge of PAABS, which
are formed due to multiple PAAR. We hope that this work can stimulate
experimental interests in S-QD-S structure, where QD may be a nanoparticle
or a gate defined geometry in 2DEG.

This project was supported by NSFC\ under Grant No. 10074001. T. H. Lin
would also like to thank the support from the State Key Laboratory for
Mesoscopic Physics in Peking University.

\smallskip $^{*}$ To whom correspondence should be addressed.

%\section* {REFERENCES}

\section*{FIGURE CAPTIONS}

\begin{itemize}
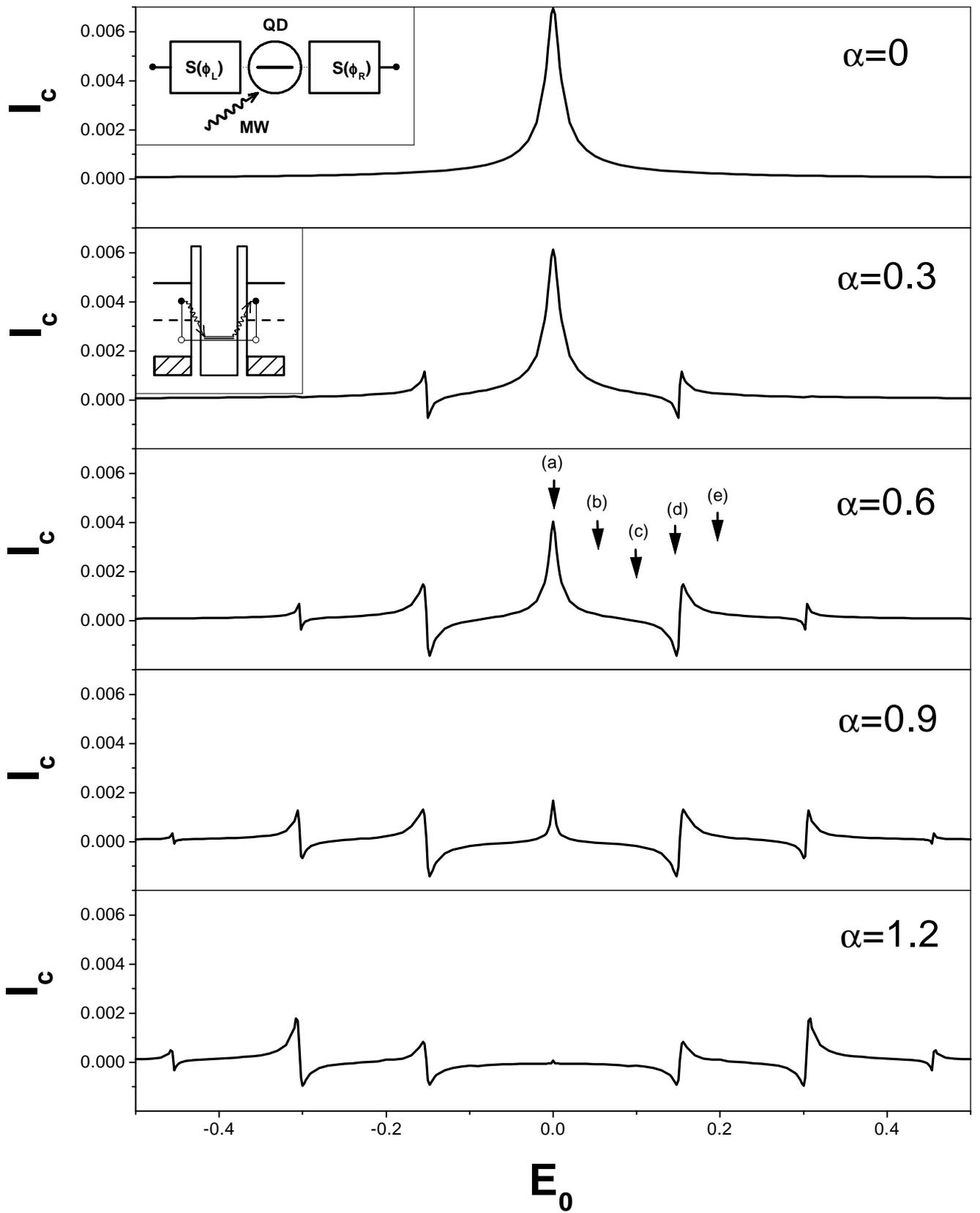
\item[{\bf Fig. 1}]  The curves of the critical supercurrent $I_c$ vs the
resonant energy level $E_0$, with the increase of MW\ strengths. Parameters
are: $\omega =0.3$, $\Gamma =0.01$. The inset in $\alpha =0$ schematically
shows the setup of an S-QD-S structure under the MW\ irradiation. The inset
in $\alpha =0.3$ demonstrates the formation of PAABS around $E_0=-\frac 12%
\hbar \omega $.

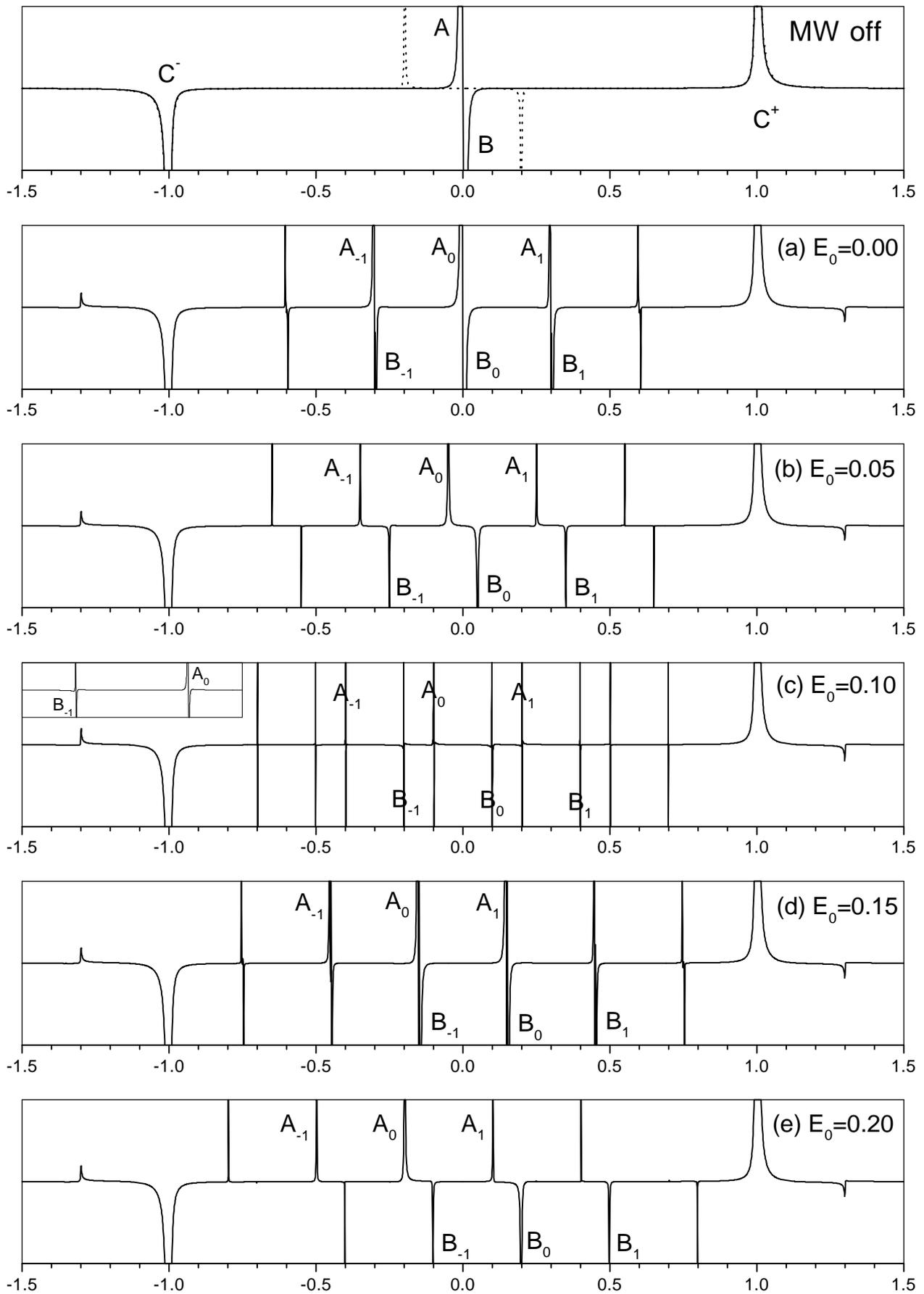
\item[{\bf Fig. 2}]  The analysis of current carrying spectrum $j(\epsilon )$
for the point (a)-(e) marked in the curves of $\alpha =0.6$ in Fig.1. For
comparison, the spectrum for $E_0=0$ (solid) and $E_0=0.2$ (dotted) without
MW irradiation are also shown on the top.

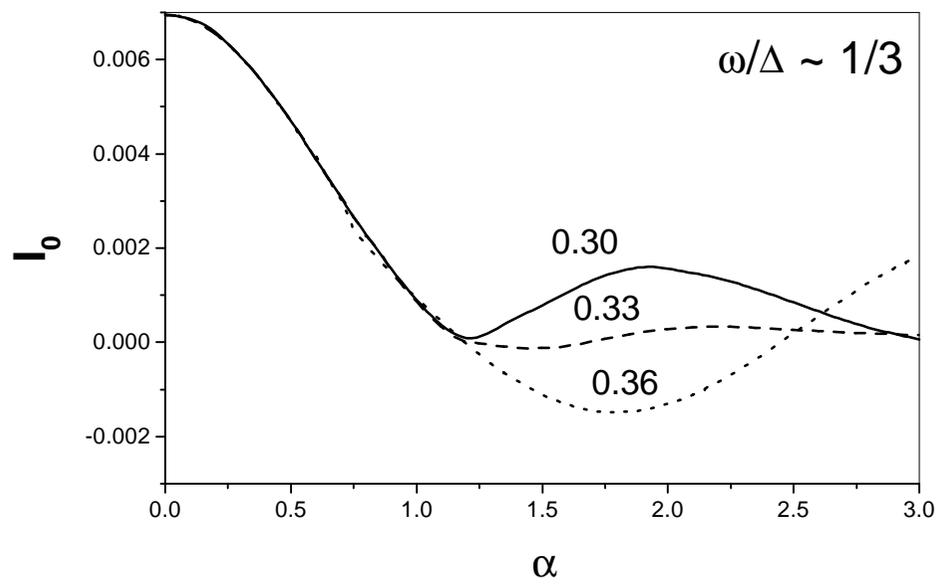
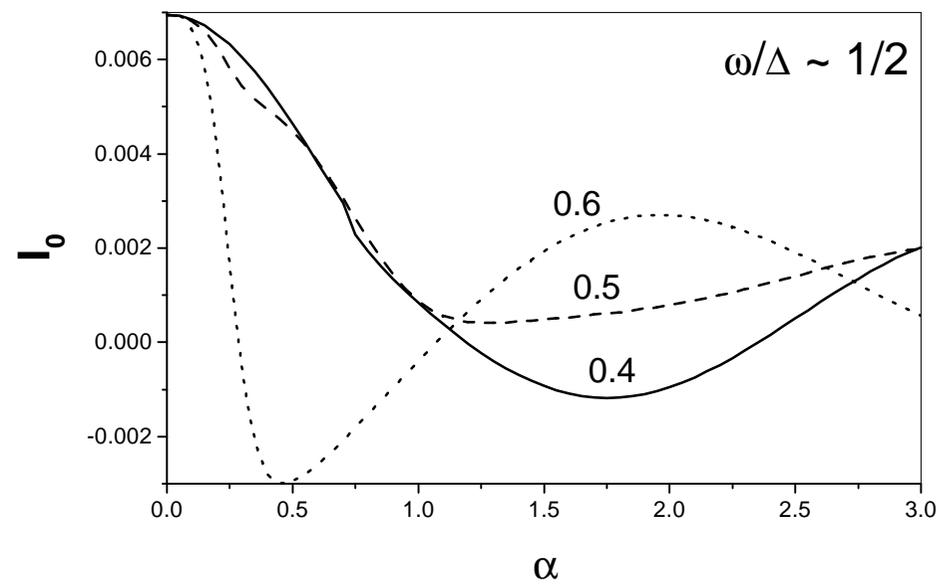
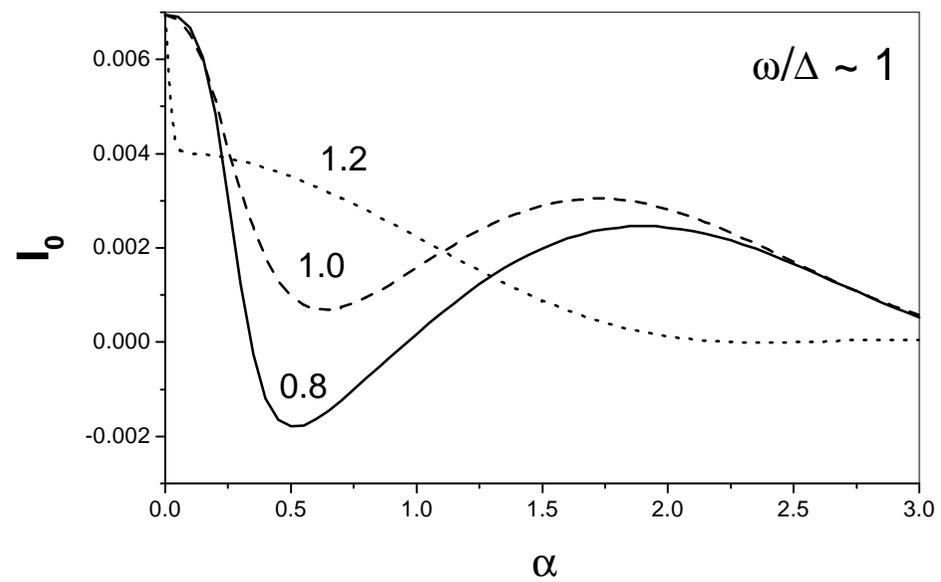
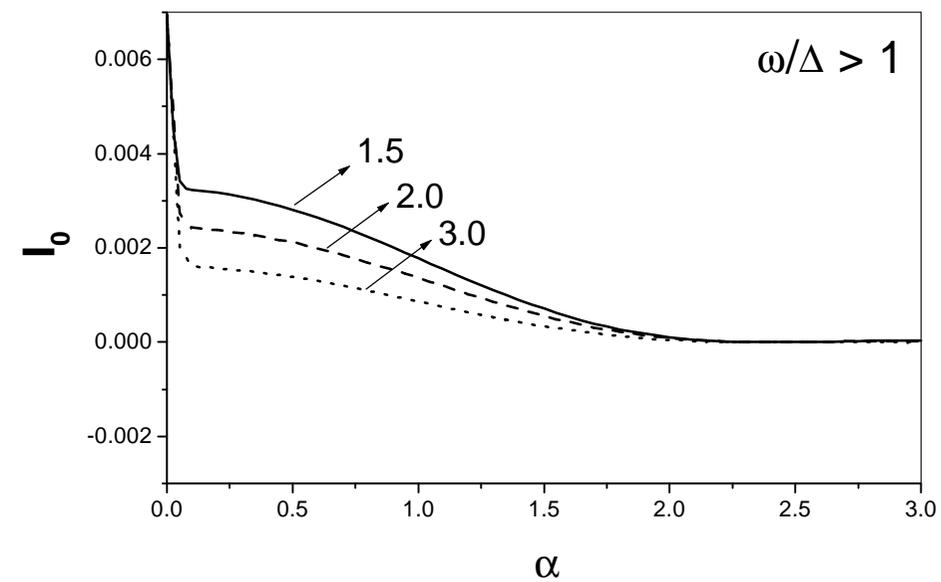
\item[{\bf Fig. 3}]  The curves of the main resonance $I_0$ vs the MW
strength $\alpha $, for MW frequencies $\omega $ chosen near $\Delta /3$,$%
\;\Delta /2$,$\;\Delta $, and $\omega >\Delta $.
\end{itemize}

\end{document}